# Charge Carrier Transport Mechanism in $Ta_2O_5$, TaON and $Ta_3N_5$ Studied by Polaron Hopping and Bandlike Models

Qianyu Zhao[a], Mengsi Cui[b], and Taifeng Liu[a]*

a) National & Local Joint Engineering Research Center for Applied Technology of Hybrid Nanomaterials, Henan University, Kaifeng 475004, China
b) Key Laboratory for Special Functional Materials of Ministry of Education, Collaborative Innovation Center of Nano Functional Materials and Applications, and School of Materials Science and Engineering, Henan University, Kaifeng, Henan 475001, China

Corresponding author: Taifeng Liu: tfliu@vip.henu.edu.cn

Abstract:

TaON and $Ta_3N_5$ are considered promising materials for photocatalytic and photoelectrochemical water splitting. But their counterpart $Ta_2O_5$ does not have good photocatalytic performance. This may cause by the different charge carrier transport mechanism in $Ta_2O_5$, TaON and $Ta_3N_5$ which are not well understood yet. Here, we investigated the charge transport mechanism in $Ta_2O_5$, TaON, and $Ta_3N_5$ by polaron and bandlike models. We find the charge transfer mechanism is small polaron hopping in $Ta_2O_5$. In TaON and $Ta_3N_5$, small polaron may not occur, and the charge may transfer through bandlike mechanism. The calculated mobility using effective mass approximation are not consistent with experimental observation. This study sets a foundation understanding of charge transport in oxynitrides and nitrides. A more accuracy model or understanding is needed to rationalize the calculated results and experimental observations in the future.

# 1. INTRODUCTION

Oxynitrides and nitrides are widely used in photocatalytic (PC) and photoelectrochemical (PEC) water splitting. This type of materials could be obtained by nitridation of its counterpart metal oxides. But oxynitrides and nitrides have smaller band gap due to the N-2p orbitals are higher than O-2p orbitals, which will cause these materials having good light absorption. The stable and commonly used phase of TaON is β-TaON (space group *P2₁/c*). The valence band maximum (VBM) and conduction band minimum (CBM) are 2.16 V and -0.34 V vs normal hydrogen electrode (NHE) at pH=0, and the band gap is 2.4 eV[1]. This type of band edge position suggests the β-TaON could oxide the water to $O_2$ and reduce $H^+$ to $H_2$ [2,3]. In the experiment, β-TaON are widely used as photoanode for water oxidation with cocatalysts[4-6], and also used as a $H_2$-evolving photocatalyst (HEP) in the photocatalytic Z-scheme overall water splitting [7]. β-$Ta_3N_5$ (space group *Cmcm*) is considered to the most important polymorphs in the practice. The VBM and CBM are 1.58 V and -0.52 V vs NHE, and the band gap is 2.1 eV[1]. β-$Ta_3N_5$ are used as particle in PC or photoanodes in PEC water splitting under visible light irradiation [4,8,9].

In photocatalytic water spitting, there are mainly three steps which are light absorption, charge separation and transport, and surface reactions. Charge separation and transport play an important role in photocatalytic water splitting. Beside good light absorption in oxynitrides and nitrides, the effect on charge carrier transfer after nitridation of its counterpart metal oxides has drawn much more attention. Using time resolved microwave conductivity (TRMC) measurements, Respinis[10] et al. observed the carrier mobility increased with the nitrogen content from $1\times 10^{-5} cm^{-2}V^{-1}s^{-1}$ in $Ta_2O_5$, to $1\times 10^{-2} cm^{-2}V^{-1}s^{-1}$ in β-TaON, until $1\times 10^{-1} cm^{-2}V^{-1}s^{-1}$ in $Ta_3N_5$. These authors suggested the charge transport mechanism in TaON are governed by small polaron hopping. Rettie[11] et al. suggested the transition-metal oxides, such as $Ta_2O_5$, the very slow charge transport are caused by the small polaron hopping. Morbec [12] et al. studied the formation of small polarons in $Ta_3N_5$ using density functional theory (DFT)+U, and they suggested the main transport mechanism for both electron and hole are bandlike. But their calculated mobility is not consistent with the experimental measurement. Respinis's[10] results unquestionably show the carrier mobility will be enhanced with high nitrogen concentration. However, the charge transport mechanism in TaON and $Ta_3N_5$ after nitridation from $Ta_2O_5$ are still not clear.

In this study, we use ε-$Ta_2O_5$, β-TaON, and β-$Ta_3N_5$ as model which are used as photocatalyst and photoanode in experiment (in the following section, simple as $Ta_2O_5$, TaON, and $Ta_3N_5$), to investigate the charge carrier transport mechanism. Polaron and bandlike models are used to gain insight of the charge transport properties. The paper is organized as following: in section 2, we give the computational details, in section 3, we present the results and discussion, and the conclusion is given in section 4.

## 2. COMPUTATIONAL DETAILS

We performed spin-polarized DFT calculations using the Vienna ab initio Simulation Package (VASP) [13,14]. For structure determination we adopted the Perdew-Burke-Ernzerh (PBE) parametrization of the exchange and correlation potential in the generalized gradient approximation (GGA)[15]. The projector-augmented wave method was applied to describing electron–ion interactions[16,17]. The structures were relaxed until all forces on atoms were less than 0.01 eV Å$^{-1}$, and a plane-wave cutoff energy of 400 eV was employed. The core electrons were represented with pseudopotentials, with 5, 6 and 11 valence electrons left in plane wave description for N, O and Ta respectively. The Γ-centered k-point mesh of the Brillouin zone sampling for the primitive cells of $Ta_2O_5$, TaON, and $Ta_3N_5$ were set at 2× 4 × 4 , 7 × 7× 7, and 5× 2× 2 based on the Monkhorst−Pack scheme [18].

We use the Heyd-Scuseria-Ernzerof (HSE) hybrid functional[19] approach, the exchanges and correction energy are given by:

$$E_{XC}^{HSE} = \alpha E_X^{SR}(\mu) + (1-\alpha)E_X^{PBE,SR}(\mu) + E_X^{PBE,LR}(\mu) + E_C^{PBE} \quad (1)$$

Where α and μ are the fraction of Hartree Fock (HF) exchange and the screening parameter, respectively. The α plays an important role to localize electron or hole to forming small polaron. Here, we keep the μ to 0.2, and only changes α. We used a 1×2×2 supercell with 112 atoms for $Ta_2O_5$, and a 2×2×2 supercell with 96 atoms for TaON, and a 3×1×1 supercell with 96 atoms for $Ta_3N_5$. For the supercells in HSE calculations, the k-points grid was reduced to the Γ point only for computational convenience.

To describe the polaron transfer process, we made use of the Marcus/Holstein two-state model, of which several descriptions exist [20-24]. The details of how to use DFT to obtain the important parameters of Marcus theory are published in several of papers [21-23,25], including the definition of an approximate reaction pathway. These papers describe also how to use the DFT data to get diffusion and mobility from Einstein's model of diffusion. Using Einstein's formula, the electron/hole mobility can be estimated approximately by

$$\mu = \frac{eD}{k_B T} = \frac{e(1-c)a^2 v_0 \exp(-\frac{\Delta G}{k_B T})}{k_B T} \quad (2)$$

where (1-$c$) is the probability that a neighboring site is available for hopping, $\Delta G$ is the energy barrier, $v_0$ is the longitudinal optical phonon frequency, $k_B$ is Boltzmann's constant, $T$ is temperature, and $a$ is the transfer distance. In this calculation, we take (1-$c$)≈1, $T$=300 K, $v_0 = 10^{13}$ Hz.

For the bandlike transport mechanism, the mobility is calculated by effective mass approximation as following:

$$\mu = \frac{2\times(2\pi)^{\frac{1}{2}}e\hbar^4 C_q}{3E_1^2(k_B T)^{\frac{3}{2}} m^{*\frac{5}{2}}} \qquad (3)$$

Where $m^*$ is the effective mass, $\frac{1}{m^*} = \frac{1}{\hbar^2}\frac{\partial^2 E(k)}{\partial k^2}$ and $E(k)$ is the energy band with k-points. $E_1$ represents the deformation potential constant of the valence-band maximum (VBM) for hole or conduction-band minimum (CBM) for electron along the transport direction, defined by $E_1^i = \Delta V_i/(\Delta l/l_0)$. Here, $\Delta V_i$ denotes the energy change of VBM and CBM when the material is compressed or dilated from the equilibrium $l_0$ by a distance of $\Delta l$. The term $C_q$ is the elastic modulus of the longitudinal strain in the propagation directions of the longitudinal acoustic wave, which can be derived from $(E-E_0)/V_0=C_q(\Delta l/l_0)^2/2$, $E$ and $E_0$ is the total energy and lattice area of the material, respectively. We use $\Delta l/l_0$ ranging from -2.0% to 2.0% to fit the values of $C_q$ and $E_1^i$ for these three materials.

## 3. RESULTS AND DISCUSSION

### 3.1 The crystal structure and polaron binding energies of $Ta_2O_5$, TaON, and $Ta_3N_5$.

The optimized crystal structures using PBE functional are shown in the Figure 1. The cell parameters of $Ta_2O_5$ are a=12.89 Å, b=4.87 Å, c=5.54 Å, α=γ=90º, β=104.3º. For TaON, they are a=5.07 Å, b=5.10 Å, c=5.25 Å, α=γ=90º, β=99.64º. For $Ta_3N_5$, the optimized lattice parameters are a=3.90 Å, b=10.31 Å, c=10.34 Å, $\alpha = \beta = \gamma = 90°$. These calculations are consistent with the experimental [8,26,27] and others calculation results[28-30].

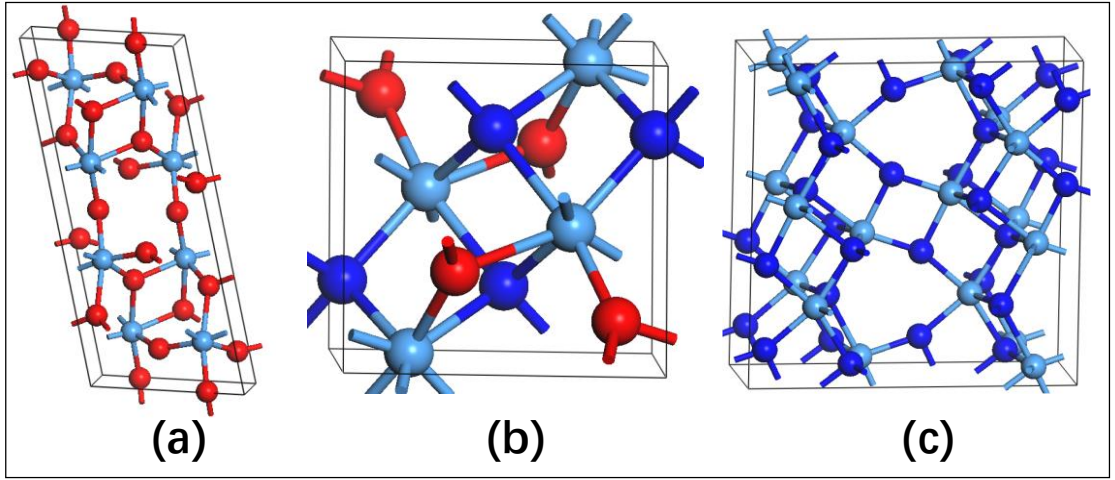

Figure 1. The crystal structures of (a) Ta$_2$O$_5$, (c) TaON, and (e) Ta$_3$N$_5$. The Cambridge blue, blue, red spheres are Ta, N, and O atoms.

Polarons are commonly exit in the metals oxidate. We use polaron binding energy to determine if polarons exit in these materials. The polaron binding energy is $E_p = \frac{m^* e^4}{8\pi^2 \hbar^2 \epsilon_{eff}^2}$, $\frac{1}{\epsilon_{eff}} = \frac{1}{\epsilon_\infty} - \frac{1}{\epsilon_s}$, where $\epsilon_\infty$ is high frequency response dielectric constant, and $\epsilon_s$ is the static dielectric constant including both ionic and electronic response. From the binding energy, it is obvious the carrier localization is favored by large effective mass and a large difference between $\epsilon_\infty$ and $\epsilon_s$ which means strong dielectric ionic response and large electron-phonon coupling. Here, we did not calculate the effective mass and dielectric constant but take them from Cui and Jiang's[28] results where they used the state-of-the-art first-principles GW approaches to calculate these parameters. The polaron binding energy in Ta$_2$O$_5$, TaON, and Ta$_3$N$_5$ are shown in Table 1. It is convenient to divide the polarons into three types by the polaron binding energies: (I) $E_p <$ 25 meV, (II) 25 $< E_p <$ 250 meV, and (III) $E_p >$ 250 meV. This separation is based on the available thermal energy ($k_BT \approx$ 25 meV) at room temperature which is related to the polaron activation energy as shown in the equation (2), and the boundary between II/III of 10 × $k_BT$ is consistent with standard approaches for classifying thermally inaccessible states[31]. Based on this criterion, we find in Ta$_2$O$_5$, the hole polaron binding energy is larger than 250 meV which indicates hole small polaron will be formed in Ta$_2$O$_5$. The electron polaron binding energy is in II area and it may not form small electron polaron in Ta$_2$O$_5$. In TaON and Ta$_3$N$_5$, the polaron binding energy are very small and in I area indicating there will no small polaron forming in these two materials.

Table 1. The polaron binding energies in Ta$_2$O$_5$, TaON, and Ta$_3$N$_5$.

| Materials | $E_p$ (hole) meV | $E_p$ (electron) meV |
|---|---|---|
| Ta$_2$O$_5$ | 404.51 | 122.64 |
| TaON | 22.81 | 16.08 |
| Ta$_3$N$_5$ | 8.13 | 10.00 |

Polaron binding energies are briefly describe the possibility of polaron forming in $Ta_2O_5$, TaON, and $Ta_3N_5$. In the following sections, we use HSE hybrid functional to realize polaronic state in these materials. An excess electron (or hole) is added or removed to the pure system, and a homogeneous background charge is assumed. Standard hybrid functional with the fraction of HF exchange α=0.25, such as PBE0, HSE06, are commonly expected to perform well for band gaps etc. in sold state. Based on this, in our case, if there are small polaron forming using hybrid functional with α around 0.25, it means the calculated results are reliable, and the small polaron may exist in this material. If small polaron state forming with α much larger than 0.25, it indicates the hybrid functional with large α over-correct the DFT error, and there may no small polaron forming in this material.

### 3.2 Electron and hole polaron transport in $Ta_2O_5$

Form the polaron binding energies in Table 1, we find hole are most likely forming small polaron in $Ta_2O_5$. We first investigate the hole polaron in $Ta_2O_5$ using HSE hybrid functional. With HSE06 functional, the hole can localize as a small polaron in $Ta_2O_5$. Hole small polaron is localized on an oxygen atom and in 2p state as shown from the spin charge density in Figure 2 (a). If consider the nearest neighbor transfer, the energy curve for hole hopping is shown in Figure 2 (b). This transfer is a diabatic process with the activation energy is 0.4 eV. In Respinis's[10] experiment, the mobility in $Ta_2O_5$ is $1.00\times10^{-5} cm^2/(Vs)$ which corresponds the activation energy of hopping is about 0.27 eV. The calculated activation energy is 0.13 eV larger than the experimental value.

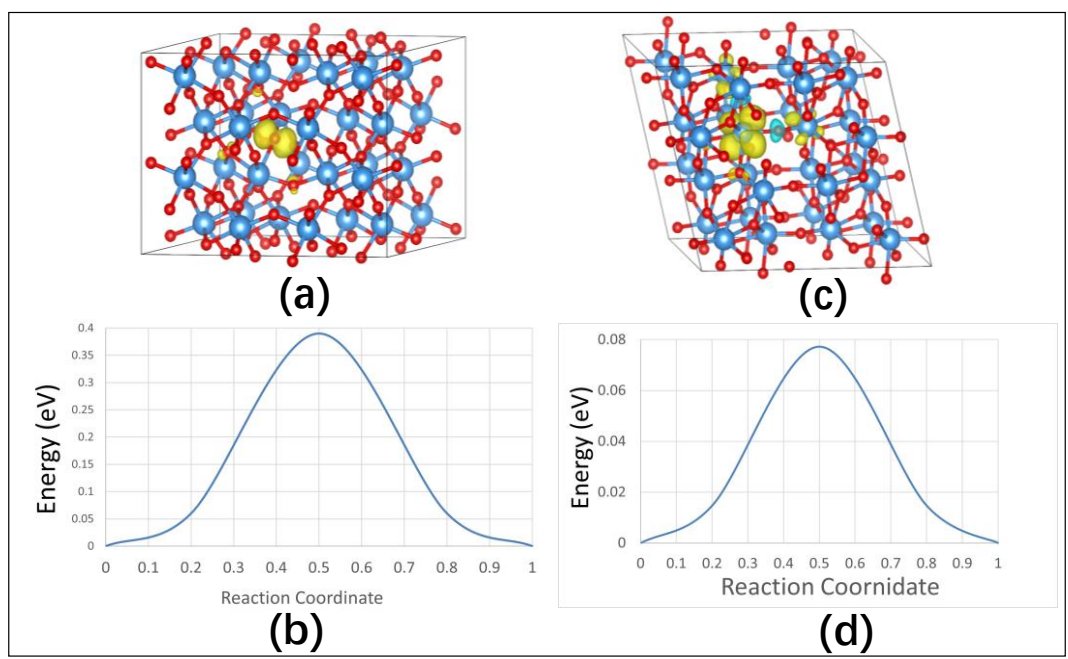

Figure 2. (a) Hole small polaron spin charge density, and (b) Hole small polaron hopping energy curves in $Ta_2O_5$; (c) Electron small polaron spin charge density, and

(d) Electron small polaron hopping energy curves in $Ta_2O_5$.

For the electron polaron, it has low possibility to form small polaron as its polaron binding energy is not too large. We find it is a delocalized state if the calculations using HSE hybrid functional with α smaller than 0.5. With α=0.5, the excess electron can localize on a Ta atom forming small polaron. The spin charge density is shown in Figure 2(c). The energy curve for the nearest neighbor hopping is shown in Figure 2 (d). The activation energy for this hopping is only 0.12 eV which is 0.15 eV smaller than the experimental value 0.27 eV. These calculations show electron small polaron in $Ta_2O_5$ may not occur.

### 3.3 Electron and Hole transport mechanism in TaON and $Ta_3N_5$

In TaON and $Ta_3N_5$, the polaron binding energies are very small and it may not be possible to form small polaron in these materials.

In TaON, we find an excess electron can be localized on a Ta atom forming small polaron using HSE hybrid functional only with α=0.6. If α is lower than 0.6, it is delocalized state. For an excess hole, it can be localized on a single oxygen atom with α larger than 0.5. But it can be localized on a single nitrogen atom forming small polaron with α=0.4. The spin charge density for electron polaron, hole polaron on an oxygen and nitrogen atoms are shown in Figure 3 (a), (b), and (c). The energy curve for hole polaron hopping between the nearest neighbor nitrogen atoms are shown in Figure 3 (d). It is an adabatic transfer with the activation energy is 0.12 eV. The calculated mobility is $1.45\times10^{-3}$ cm$^2$/(Vs) which is about 10 times smaller than the experimental measurement with a value $1.00\times10^{-2}$ cm$^2$/(Vs) by TMRC[10]. Based on these findings, we consider hole small polaron on nitrogen atoms may occur in TaON. For the electron polaron and hole polaron on oxygen atoms, all the calculations do not support its forming small polarons in TaON.

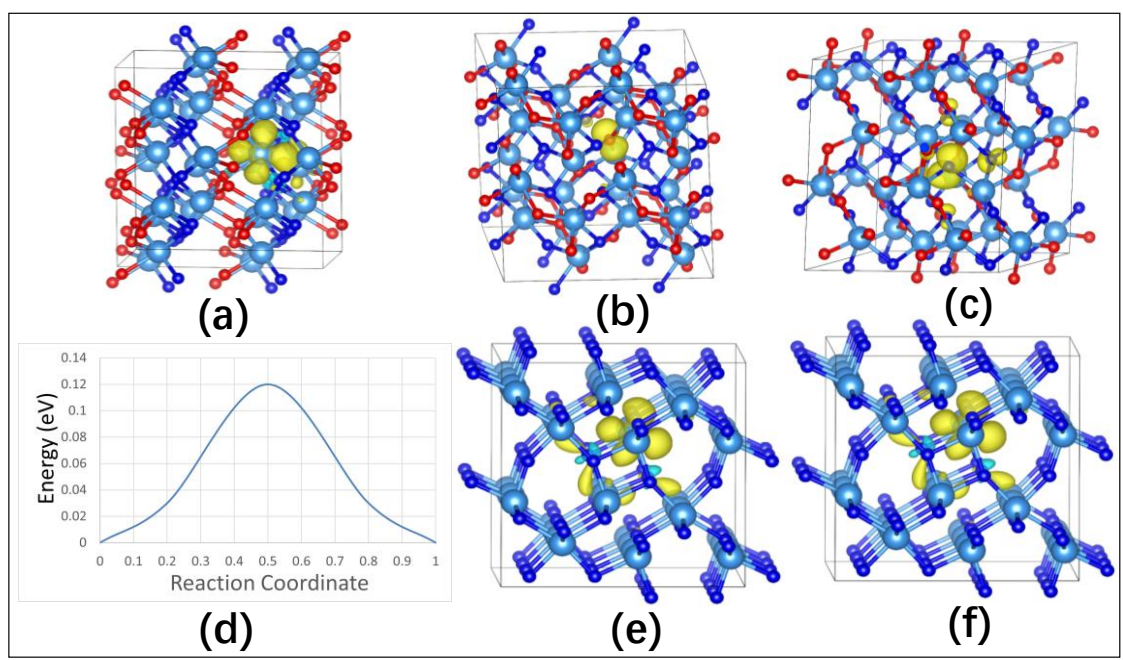

Figure 3. Spin charge density for (a) electron polaron, hole polaron on an (b) oxygen and (c) nitrogen atoms in TaON; (d) energy curve for hole polaron on nitrogen atom hopping in TaON; (e) spin charge density for electron polaron and (f) hole polaron in $Ta_3N_5$.

In $Ta_3N_5$, an excess electron can be localized on a single Ta atom using hybrid functional with α larger than 0.7. For an excess hole, it is delocalized state among all nitrogen atoms with α smaller than 0.6. The spin charge density is shown in Figure 3 (e) and (f). Polarons do not occur, and the charge transport mechanism maybe bandlike in $Ta_3N_5$ which Morboc [12] et al. obtained the same conclusion.

3.4 Bandlike transport mechanism for electron and hole in $Ta_2O_5$, TaON, and $Ta_3N_5$

In this section, we use the effective mass approximation equation (3) to investigate the bandlike charge transport in $Ta_2O_5$, TaON, and $Ta_3N_5$. The effective mass is taken from Cui and Jiang's [28] results. We calculated deformation potential constant $E_l$, elastic modulus $C_q$ using PBE functional. The calculated mobility μ and other parameters are shown in Table 2. The calculated $C_q$ and $E_l$ of are shown in Figure 4. In $Ta_2O_5$, the calculated mobility using effective mass approximation for hole and electron are 2.476 and 22.535 $cm^2$/(Vs). But in experiment, it is $1.00 \times 10^{-5} cm^2$/(Vs). That is because in $Ta_2O_5$, the carrier transport is small polaron hopping not bandlike model. For the TaON, the calculated mobility is about hundreds, but in the experiment, the value is $1.00 \times 10^{-2} cm^2$/(Vs), which is about four orders of magnitude larger. For the $Ta_3N_5$, the mobility reaches thousands while in the experiment is $1.00 \times 10^{-1} cm^2$/(Vs), which is still four orders of magnitude larger. From this point, we did not rationalize the calculated results and experimental observed values. The probable reason may be in experiment there are defects in the materials, but in our calculations, we only consider the pure structures.

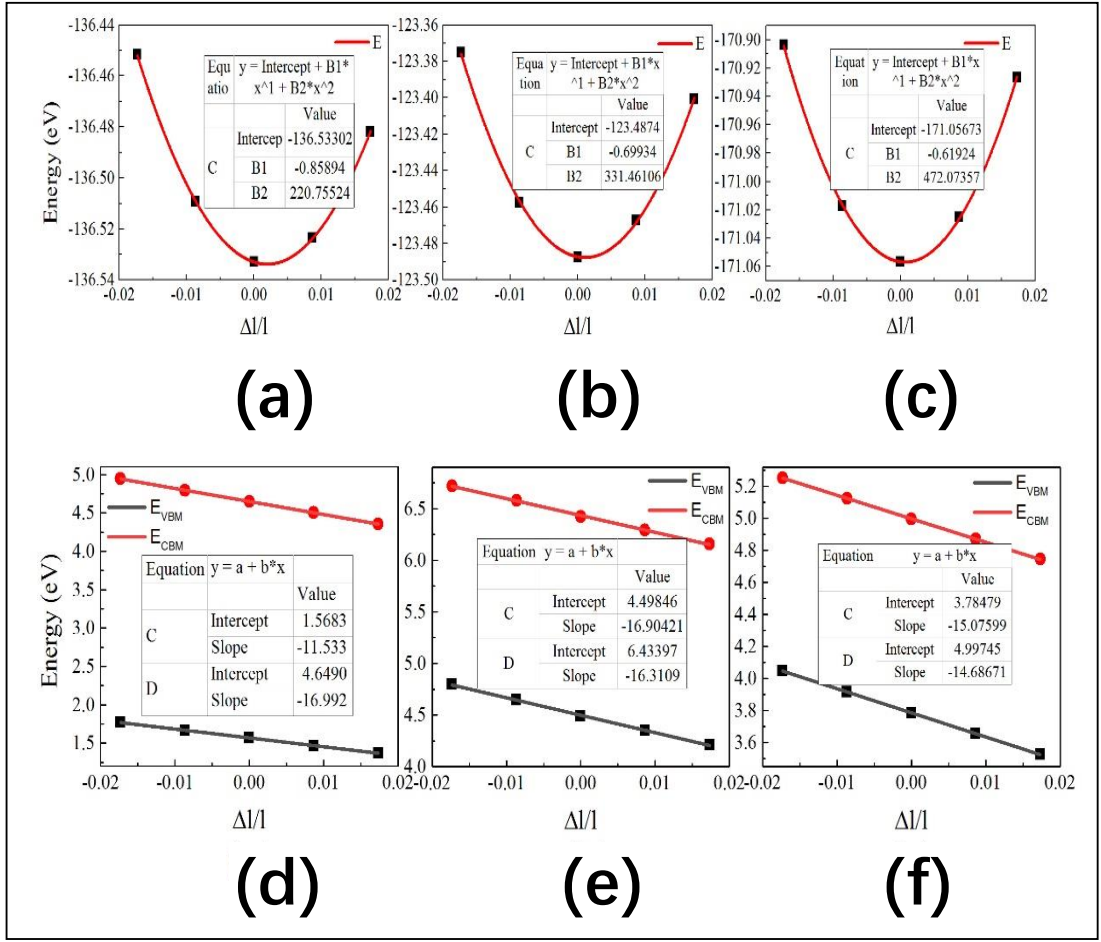

Figure 4. The elastic modulus Cq of (a) $Ta_2O_5$, (b) TaON, and (c) $Ta_3N_5$, and the deformation potential constants $E_1$ of (d) $Ta_2O_5$, (e) TaON, and (f) $Ta_3N_5$.

Table 2. Deformation potential constant $E_1$, elastic modulus $C_q$, and calculated mobility μ using effective mass approximation in $Ta_2O_5$, TaON, and $Ta_3N_5$.

| Materials | Carrier type | $E_1$(eV) | $C_q$ ($10^{11}Jm^{-3}$) | μ ($cm^2V^{-1}s^{-1}$) |
|---|---|---|---|---|
| $Ta_2O_5$ | h | -11.534 | 4.072 | 2.476 |
|  | e | -16.993 |  | 22.535 |
| TaON | h | -16.904 | 8.120 | 105.950 |
|  | e | -16.311 |  | 272.398 |
| $Ta_3N_5$ | h | -15.076 | 7.253 | 10000.052 |
|  | e | -14.687 |  | 626.379 |

4. **CONCLUSIONS**

In this work, the charge transport mechanism in $Ta_2O_5$, TaON, and $Ta_3N_5$ were investigated by polaron and bandlike models. The polaron binding energies show hole polaron in $Ta_2O_5$ occurs but polarons in TaON and $Ta_3N_5$ may not occur. Hole small polaron in $Ta_2O_5$ can be obtained using HSE06 functional while electron small polaron can only appear using HSE hybrid functional with α larger than 0.5. In TaON and $Ta_3N_5$, carriers can be localized to form small polaron using HSE hybrid functional with α larger than 0.5 except hole small polaron on nitrogen atom in TaON which with α=0.4 is enough to form small hole polaron. Large fraction of HF in HSE hybrid functional can over-correct DFT error which indicates small polaron may not occur in TaON and $Ta_3N_5$. The calculated mobility using effective mass approximation is at least three order magnitudes larger than the experimental measurement in TaON and $Ta_3N_5$. This study sets a foundation understanding of charge transport in oxynitrides and nitrides. In further work, maybe other models, are needed to rationalize the calculated results and experimental observation.

**AUTHOR INFORMATION**
*Corresponding Author:*
Taifeng Liu: tfliu@vip.henu.edu.cn

**Notes**
The authors declare no competing financial interest.

ACKNOWLEDGMENT
This work was supported by the National Natural Science Foundation of China (grant # 21703054).